\documentclass[aps,prl,reprint,showpacs,superscriptaddress]{revtex4-1}
\usepackage{graphicx}
\usepackage{amsfonts}
\usepackage{txfonts} 
\usepackage{times}
\usepackage{subfigure}
\usepackage{textcomp}
\usepackage{color}
\definecolor{gray}{gray}{0.8}
\newcommand{\beq}{\begin{equation}}     \newcommand{\eeq}{\end{equation}}
\newcommand{\beqa}{\begin{eqnarray}}    \newcommand{\eeqa}{\end{eqnarray}}
\newcommand{\bde}{\begin{description}}  \newcommand{\ede}{\end{description}}
\newcommand{\ben}{\begin{enumerate}}    \newcommand{\een}{\end{enumerate}}

\newcommand{\la}{\langle}               \newcommand{\ra}{\rangle}

\newcommand{\kT}{{k_{\rm B}T} }

\newcommand{\eqn}[1]{\beq{ #1 }\eeq}

\newcommand{\inv}[1]{{\frac{1}{#1}}}

\newcommand{\inRbracket}[1]{{\left({#1}\right)}}

\newcommand{\rev}[1]{{#1}}

\begin{document}
\title{Momentum transfer in non-equilibrium steady states 
}

 \author{Antoine Fruleux} \affiliation{Mati\`{e}res et Syst\`{e}mes Complexes, CNRS-UMR7057, Universit\'e   Paris-Diderot, 75205 Paris, France} \affiliation{Gulliver, CNRS-UMR7083, ESPCI, 75231 Paris, France} 
 \author{Ryoichi Kawai} \affiliation{Department of Physics, University of Alabama at Birmingham, Birmingham, AL 35294, USA}
 \author{Ken Sekimoto} \affiliation{Mati\`{e}res et Syst\`{e}mes Complexes, CNRS-UMR7057, Universit\'e   Paris-Diderot, 75205 Paris, France}  \affiliation{Gulliver, CNRS-UMR7083, ESPCI, 75231 Paris, France}

\begin{abstract}
When a Brownian object interacts with non-interacting gas particles under non-equilibrium conditions,
the energy dissipation associated to the Brownian motion causes an additional force on the object as a `momentum transfer deficit'.  This principle is demonstrated first by a new NESS model and then applied to several known models such as adiabatic piston for which simple explanation has been lacking.
\end{abstract}

\date{\today  
}
\maketitle

In nonequilibrium statistical mechanics the {mechanical} coupling between a system and environments still
remains poorly understood.
In the Langevin description, the framework of energetics was developed during the last decade\cite{sekimoto97,LNP} but there are certainly many aspects which cannot be grasped by such a level of description. 
For example, when a Brownian object is not symmetric, such as a cone or wedge shape, its asymmetric properties are
not fully reflected in the linear friction constant or tensor, $\gamma$, of the Langevin equation because $\gamma$ is non-polar.

Related to this limitation, or due to our lack of {comprehension} about nonequilibrium Brownian motion, there are a class of nonequilibrium phenomena which have refused to be understood at a fundamental level. 
An interesting example is the {\it adiabatic piston} separating two gases of different temperatures under pressure equilibrium \cite{Callen2,feynman63-ch39,Piasecki1999463}. 
The laws of thermodynamics cannot tell whether the piston moves or not \cite{Callen2}. 
Feynman \cite{feynman63-ch39} 
pointed out that the fluctuations of piston's velocity should be taken into account.
However,  the Langevin description with linear friction falsely predicts zero mean velocity.
The adiabatic piston is, therefore, still listed among major unsolved thermodynamics problems \cite{Lieb1999}.  
This difficulty is also shared by some models of Brownian ratchet motors working between ideal gas reservoirs \cite{VdB-prl2004-ratchet,*VdB-njp2005-ratchet}.

A common solution to these problems is to resort to full and general microscopic descriptions, such as the molecular dynamic (MD) simulation or master-Boltzmann equation under pertinent perturbative approximations \cite{Piston-Lebowitz-PR1959}.
These methods are effective in predicting the outcome. For the adiabatic piston, the MD simulations \cite{vdB-adpiston-EPL2000} and perturbative master-Boltzmann equation \cite{vdB-adpiston-EPL2000,Piasecki1999463,adiab-piston-infini-Gruber1999,Gruber-PhysicaA1999,*Gruber2003} give quite consistent results showing that the piston moves towards the \textit{hotter} reservoir. 
For the ratchet models the agreement between MD simulation and perturbative theory is excellent \cite{VdB-prl2004-ratchet,*VdB-njp2005-ratchet}. When higher order terms are taken into account, the perturbative theories can tell the effect of the shape of Brownian object \cite{VdB-prl2004-ratchet,*VdB-njp2005-ratchet} or of the inelasticity of collisions, called {\it inelastic piston}  \cite{0295-5075-82-5-50008,PhysRevE.82.011135} and their combinations \cite{PhysRevE.75.061124,ineltri-VdB-EPL07}.
Yet,  we still have no physical explanation why the nonequilibrium processes give rise to a force and what determines its direction.

In this paper, we will develop a general theoretical framework to answer to this fundamental problem. The key is to explicitly take into account the momentum and mass balances under nonequilibrium condition, in addition to the energy balance considered by the stochastic energetics \cite{LNP}. Briefly, the nonequilibrium energy transfer, or dissipation, leads to a deficiency in the momentum transfer from the environment to the Brownian object, while the gas particle (mass) flux is unchanged by the dissipation. We shall call this deficiency 
the {\it momentum transfer deficit due to dissipation} or {MDD}, for short. 
We will show that this MDD is expressed as
a form of nonequilibrium boundary condition for the momentum flow [Eq.~(\ref{eq:Jp}) below]. 
With this condition, many nonequilibrium problems which have been hitherto solved case-by-case can be explained in a unified manner sometimes even at semi-quantitative level. 

 In the following, we first describe the basic principle. To demonstrate the principle we introduce a simple model of nonequilibrium steady state (NESS) and its solution. 
 Then, we will apply the basic principle to unexplained problems such as adiabatic piston in order to show the universality of underlying physics. We extend our principle to include weak inelasticity of collisions between the Brownian object and gas particles.

\begin{figure}[bt]{}
\centering{
\includegraphics[width=3.2in]{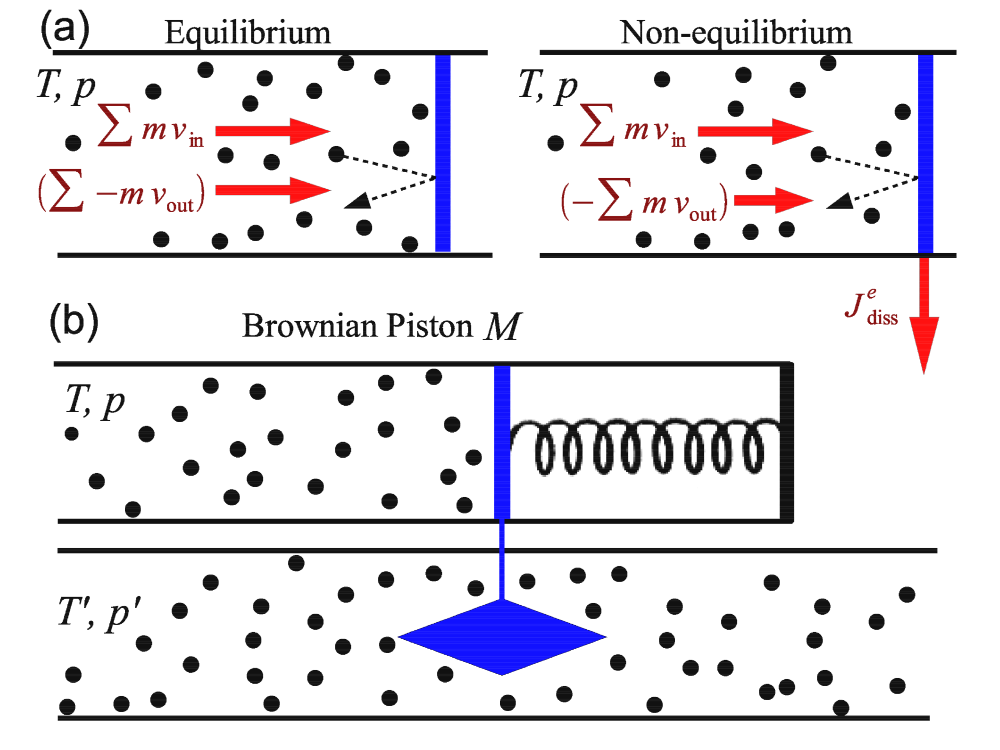}
}
\caption{(a) When the piston's mean velocity $\bar{V}$ is zero,
the net momentum transfer by the incoming particles, $\sum mv_{\rm in},$  and by the outgoing particles, $\sum (-mv_{\rm out}),$ per unit time sum up to give the force on the piston.
({\it Left}) Equilibrium state where no net energy is transferred to the piston. 
({\it Right}) 
Non-equilibrium case where the energy is dissipated at a rate $J_{\rm diss}^{(e)}$.  
When $J_{\rm diss}^{(e)}>0 [<0]$, additional force $F_{\it MDD}<0 [>0]$ is exerted on the piston.~~
(b) Cooled/warmed Brownian piston : The piston (thick bar) and object (diamond) are tightly bound and are held by a spring. The gas environments have temperatures $T$ and $T'$ and pressures $p$ and $p'$, respectively.
\label{fig:1}} 
\end{figure}

In the elementary setup Fig.~\ref{fig:1}(a), an ideal gas of temperature $T$ and pressure $p$ fills to the left of the wall. The wall is a Brownian object and  its velocity fluctuates.  However, it macroscopically remains at rest. The collisions of the gas particles with the wall strictly satisfy the momentum conservation but can be either elastic or inelastic.  
We assume that the energy transfer by individual collision is very small so that 
the double-collision by the wall with the same gas particle is negligible.
More specifically, the mass of the wall, $M$, and that of gas particles, $m$, are assumed to satisfy $\epsilon^2\equiv m/M\ll 1.$
Our interest is the force exerted on the wall by the gas or equivalently the momentum transfer from the gas to the wall.
We separate this momentum transfer into two parts; one due to the incoming particles toward the wall, $\sum mv_{\rm in}$, and the other to the outgoing particles from the wall, $\sum (-mv_{\rm out}$), where the sum is taken over a unit time. The sum of the two momentum fluxes gives the force on the wall.

When the wall's microscopic fluctuations are thermally in equilibrium with the gas [Fig.~\ref{fig:1}(a)(left)], the detailed balance condition tells that two momentum fluxes should be equal on the time average, and the sum of the two is the hydrostatic pressure $p$ times the surface area, $L$. Note that unlike a simple kinetic theory used in elementary textbooks the individual collisions can transfer energy between the gas and the wall at the microscopic level since the wall fluctuates. It is the detailed balance that makes the two momentum fluxes identical on the average.
On the other hand, when the dissipation carries away a part of kinetic energy of the gas upon collision to the outside of the system at the rate  $J^{(e)}_{\rm diss}$ per unit time
[Fig.~\ref{fig:1}(a)(right)], the speed of the outgoing particles is, on the average, less than that of the incoming ones. Therefore, the momentum transfer by outgoing flux, $\sum (-mv_{\rm out})$, should be less than the incoming one,  $\sum mv_{\rm in}$, which should not be influenced by the dissipation as long as the double collisions are negligible. This reduction in momentum transfer is the {MDD}, and the resulting force on the wall is less than that in the equilibrium by the {MDD}. This additional force due to {MDD} is exactly the point where the Langevin equation with linear friction fails to grasp the left-right asymmetry of the system.

To make this principle more concrete and quantitative, we first assume elastic collisions between gas particles and the wall. We take the thermal velocity $v_{\rm th}=\sqrt{\kT/m}$ as a typical normal component of the velocity of incoming particles $v_{\rm in}$ up to a numerical factor (see below).
The first part of momentum transfer is  $\sum mv_{\rm in}\simeq mv_{\rm th}\,{\omega_{\rm col}}$, where 
\rev{$\omega_{\rm col}\simeq \rho Lv_{\rm th}/2$ is} the collision frequency on the wall. 
We denote by $v'(<0)$ the typical normal component of the outgoing velocities $v_{\rm out}$. 
The second part of momentum transfer is then $\sum (-mv_{\rm out})\simeq m|v'|{\omega_{\rm col}}.$ 
The conservation of mass fluxes imposed the common frequency, $\omega_{\rm col}$ for both incoming and outgoing fluxes.

Now $v'$ is related to $v_{\rm th}$ though the energy balance condition, 
$\frac{m}{2}{v_{\rm th}}^2-\frac{m}{2}{v'}^2\simeq \frac{J^{(e)}_{\rm diss}}{\omega_{\rm col}}.$
Here, we assumed that the parallel component of the velocity does not contribute to the energy loss.
Noting $|v'|\simeq v_{\rm th}$ for weak energy transfer, the left hand side can be approximated by $v_{\rm th}(mv_{\rm th}-m|v'|)$. Then the {MDD} par unit time is $(mv_{\rm th}-m|v'|)\omega_{\rm col}\simeq \frac{J^{(e)}_{\rm diss}}{v_{\rm th}}$, and the net force on the wall is
\eqn{\label{eq:Jp}
F=F_{\rm eq}+ F_{\rm {MDD}}, \quad F_{\rm {MDD}}\simeq-c \frac{J_{\rm diss}^{(e)}}{v_{\rm th}}, }  
where $F_{\rm eq}=pL$ is equilibrium hydrostatic force and the numerical constant $c$ is 1 in 
the above semi-quantitative derivation. 
From the view of the gas, Eq.~(\ref{eq:Jp}) can be considered as a boundary condition for the momentum flux.
This additional force $F_{\rm MDD}$ induced by dissipation is the main result of the present Letter.

An interesting realization of {MDD}, which is also a new model of NESS, is illustrated in Fig.~\ref{fig:1}(b). In two dimension, a piston with smooth vertical wall as Brownian object is in contact with a gas of temperature $T$ and pressure $p$. Its horizontal motion is tightly coupled to another object (rhombus) immersed in a different gas environment of temperature $T'$. We can show that when the horizontal diagonal $\ell_\parallel$ and vertical diagonal $\ell_\perp$ of the rhombus are made indefinitely large keeping $\ell_1 \equiv 2\ell_\perp^2/\ell_\parallel$ constant, the collisional forces from the second gas converges to the ordinary force of Langevin equation, that is the frictional, $-\gamma' V$ and random force, $\sqrt{2\gamma'\kT'}\Theta_t$, with the friction constant $\gamma'=\sqrt{\pi/8} \,\rho \ell_1 m v_{\rm th}$ ($\rho=p/\kT$) and Gaussian white noise $\Theta_t$ with $\la \Theta_t\Theta_{t'}\ra=\delta(t-t')$ \footnote{We will discuss the details somewhere else.}. 
Therefore, for $T'<T$ the thermal contact dissipates energy to the second gas without net transport of momentum between the two gases on the time average.

For weak dissipation, the dissipation rate  $J^{(e)}_{\rm diss}$ depends on the coupling with the environments only through the friction constants, $\gamma$ with the first gas and the aforementioned $\gamma'$ with the second gas: 
\eqn{\label{eq:Je}
J_{\rm diss}^{(e)}=\frac{\kT-\kT'}{M(\gamma^{-1}+\gamma'^{-1})}.}
\rev{For later use, we here show a heuristic derivation of Eq.~(\ref{eq:Je}).}
 Assuming a kinetic temperature of Brownian motion, $\kT_{\rm kin}$,  we can construct dimensionally $J_{\rm diss}^{(e)}$ by the time constant $M/\gamma$ and the temperature gap, $\kT-\kT_{\rm kin}, $ as $J_{\rm diss}^{(e)}=(\gamma/M) (\kT-\kT_{\rm kin})$. Applying the same argument for the second bath, i.e. $J_{\rm diss}^{(e)}=- (\gamma'/M) (\kT'-\kT_{\rm kin})$, and eliminating $T_{\rm kin}$, we obtain Eq. (\ref{eq:Je}). 
\rev{ For the standard derivation, see Ref.~\cite{LNP}.
The linear friction, $\gamma= c'\rho L mv_{\rm th},$ 
can be also obtained heuristically by a Doppler shift of the momentum transfer, 
where $c'=\sqrt{8/\pi}$ from the standard gas kinetics.}

Now substituting Eq. (\ref{eq:Je}) to (\ref{eq:Jp}) we have a concrete form of {MDD} and the force in nonequilibrium. We can show that the microscopic approach with the master-Boltzmann equation gives exactly the same result if we choose $c=\sqrt{\pi/8}$. When the wall is `cooled', i.e. for $T'<T$, the mean position of the wall in NESS is displaced leftwards relative to its equilibrium position \rev{, and {\it vice versa}}.

The principle (\ref{eq:Je}) is also applicable to the case where the collision is weakly inelastic. In Fig.~\ref{fig:1}(b) we remove the rhombus and the second gas environment, and instead assume the 
restitution coefficient  $e$ ($1-e\ll 1$) for the collision between the gas particles and the vertical wall. 
In this case, the dissipation rate consists of two parts: 
\eqn{J^{(e)}_{\rm diss}=J^{(e)}_{\rm diss, hk}+J^{(e)}_{\rm diss, ex}.}
The `house-keeping' heat generation \cite{Oono-Paniconi98}, $J^{(e)}_{\rm diss, hk},$ 
is due to inelasticity of individual collisions. The `excess' dissipation, $J^{(e)}_{\rm diss, ex},$ is due intrinsically to the nonequilibrium Brownian motion of the wall. 
If the wall were rigidly fixed, only $J^{(e)}_{\rm diss, hk}$ is nonzero. In this case 
the dissipation per collision is $mv_{\rm th}^2/2-m{v'}^2/2=(1-e^2)mv_{\rm th}^2/2$ and 
 $J^{(e)}_{\rm diss, hk}=(1-e^2)mv_{\rm th}^2/2\times \omega_{\rm col}$.
\rev{Noting $(1-e)^2\simeq 2(1-e)$,} the same argument leading to Eq.~(\ref{eq:Jp}) gives
\eqn{\label{eq:Jpe}
F_{\rm {MDD}}\simeq F_{\rm {MDD},hk} - c \frac{J_{\rm diss,ex}^{(e)}}{v_{\rm th}}, 
\quad F_{\rm {MDD},hk}=-\frac{1-e}{2} pL,}  
where $F_{\rm {MDD},hk}$  is force due to the `house-keeping' {MDD} 
which reduces the force even for a fixed wall. (A sand bag will receive less impact than a hard wall by a bullet.)

The excess dissipation is expressed in terms of the aforementioned kinetic temperature $\kT_{\rm kin}$ as $J^{(e)}_{\rm diss,ex}=(M/\gamma)(\kT-\kT_{\rm kin})$. 
Upon a binary inelastic collision,
the velocity of a Brownian object changes in the same way as that of elastic collision if the effective mass
$M_{\rm eff}\equiv 2M/(1+e)$ is used.
The Brownian object then obeys approximately the Maxwell distribution $\propto e^{-M_{\rm eff} V^2/(2\kT)}$. It implies $\kT_{\rm kin}=\kT\times (1+e)/2.$ Therefore,
\eqn{\label{eq:Jeex}
J^{(e)}_{\rm diss,ex}=\frac{\gamma}{M}\frac{1-e}{2}\kT,}
where the friction constant $\gamma$ is the same as before in the lowest order in $(1-e)$.
With the numerical factor $c=\sqrt{\pi/8}$, we recover the microscopic result. 
When the dominant house-keeping {MDD} is canceled by the same {MDD} from the other sides, as for an
inelastic triangular Brownian object \cite{PhysRevE.75.061124}, it is the excess {MDD} that explains the origin of nonequilibrium force. 

We have shown that our simple calculation gives the identical result as microscopic approaches up to a numerical factor of order one. We note that the microscopic approach is still needed to find the correct numerical factor for the Brownian object of complicated geometry and to find higher order corrections to the perturbation. However, our main goal is rather to show that the principle (\ref{eq:Jp}) is a general theory of the force under nonequilibrium condition. Below we will apply the basic schema Fig.~\ref{fig:1} to various known cases and show how our principle based on the conserved quantities is fundamental to understand the phenomena. 
\begin{figure}[b]{}
\centering{
\includegraphics[width=2.9in]{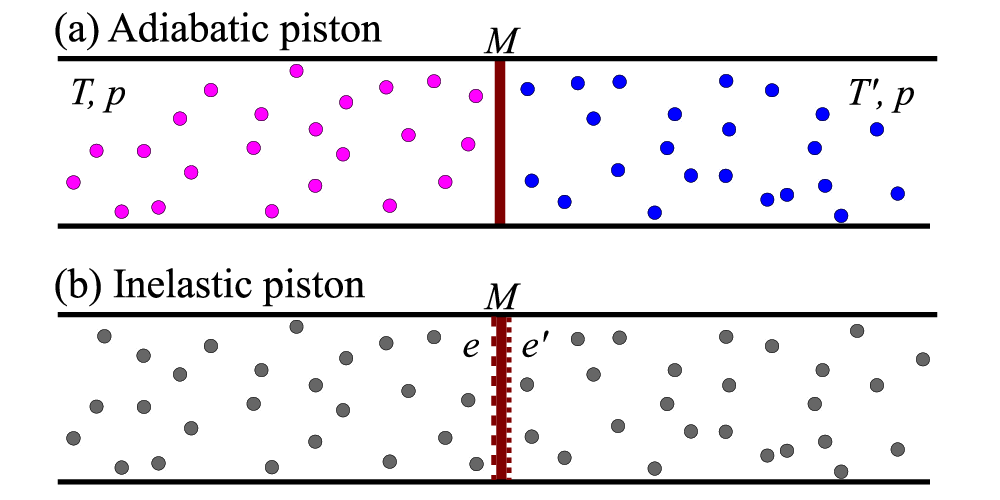}
}
\caption{(a) A microscopic ``adiabatic'' piston of mass $M$ (vertical bar) separates two semi-infinite gases of point-like particles with mass $m(\ll{}M)$. The two gases have the same pressure, $p,$ but different temperatures, $T$ and $T'$. 
 (b) A macroscopic inelastic piston with restitution coefficients, $e$ (left surface) and $e'$ (right surface) 
is in a gas. \label{fig:2}} 
\end{figure}%

{\it Adiabatic piston (with elastic wall)} \cite{Piasecki1999463,adiab-piston-infini-Gruber1999,Gruber-PhysicaA1999,vdB-adpiston-EPL2000} : 
We apply the boundary conditions (\ref{eq:Jp}) to the both sides of the piston shown in Fig ~\ref{fig:2}(a), with an appropriate sign of the forces and dissipation rates as well as taking account of different temperatures. By the isobaric condition, the equilibrium force $F_{\rm eq}$ on both sides cancels. On the side of temperature $T'$  the force, $F'_{\rm {MDD}},$ should contain the dissipation rate $J^{(e)\prime}_{\rm diss}= -J^{(e)}_{\rm diss}$ to assure the energy conservation. Both $F_{\rm {MDD}}$ and $F'_{\rm {MDD}}$ are oriented toward hot side (leftward if $T>T'$), and the total momentum balance about the piston is recovered by the frictional force, $F_{\rm {MDD}}+F'_{\rm {MDD}}-(\gamma+\gamma')\bar{V}=0,$ where $\bar{V}$ is the steady state velocity of the piston. Combining with Eqs.~(\ref{eq:Jp}) and (\ref{eq:Je}) the steady state velocity reads
\eqn{\label{eq:Vbar}
\bar{V}=-\frac{c}{\gamma+\gamma'}\inRbracket{\inv{v_{\rm th}}+\inv{v'_{\rm th}}}
\frac{\kT-\kT'}{M(\gamma^{-1}+\gamma'^{-1})},}
which is identical to the result obtained from the perturbative calculation in \cite{adiab-piston-infini-Gruber1999} with $c=\sqrt{\pi/8}$. ($p=\rho \kT$ is assumed to verify it.)
 The correction to dissipation due to the `mesoscopic loss' $(\gamma+\gamma')\bar{V}^2$ is of higher order by $\epsilon^2$ and, therefore, negligible. This remark applies to all other examples below.

{\it Inelastic piston} \cite{0295-5075-82-5-50008,PhysRevE.82.011135} :  
A piston of two inelastic faces shown in Fig.~\ref{fig:2}(b)  is in a gas of temperature $T$ and pressure $p$, and the faces have coefficients of restitution $e$ (left face) and $e'$ (right face), respectively, with 
$1-e\ll 1$ and $1-e'\ll 1$. The dissipation rate  $J^{(e)}_{\rm diss}$, {MDD} and $F_{\rm {MDD}}$, on each face satisfy Eq.~(\ref{eq:Jp}). But the {MDD} on the two faces has different signs, and thus the net force arises only when $e\neq e'$,  as $\left.F_{\rm {MDD}}\big|_{\rm left}+\right.F_{\rm {MDD}}\big|_{\rm right}=(e-e')pL/2$. 
By balancing with the frictional force, the stationary velocity $\bar{V}$
to the lowest order in $1-e$ and $1-e'$ (therefore $|e-e'|\ll 1$) and in $\epsilon$  is 
\eqn{\label{eq:VinelP}
\bar{V}=\frac{1}{\gamma+\gamma'}\frac{e-e'}{2}pL = -\frac{e-e'}{4c'}v_{\rm th},} 
where $\gamma\simeq \gamma'=c'\rho L mv_{\rm th}$ with $c'$ a constant.
The result (\ref{eq:VinelP}) agrees with the perturbative results 
\cite{0295-5075-82-5-50008,PhysRevE.82.011135}  with $c'=\sqrt{8/\pi}$.
This elementary example shows, however, that our principal formula (\ref{eq:Jp}) is universal whether or not the origin of dissipation is kinematical or dynamical, because the momentum conservation is universally valid.

\begin{figure}[bt]{}
\centering{
\includegraphics[width=2.9in]{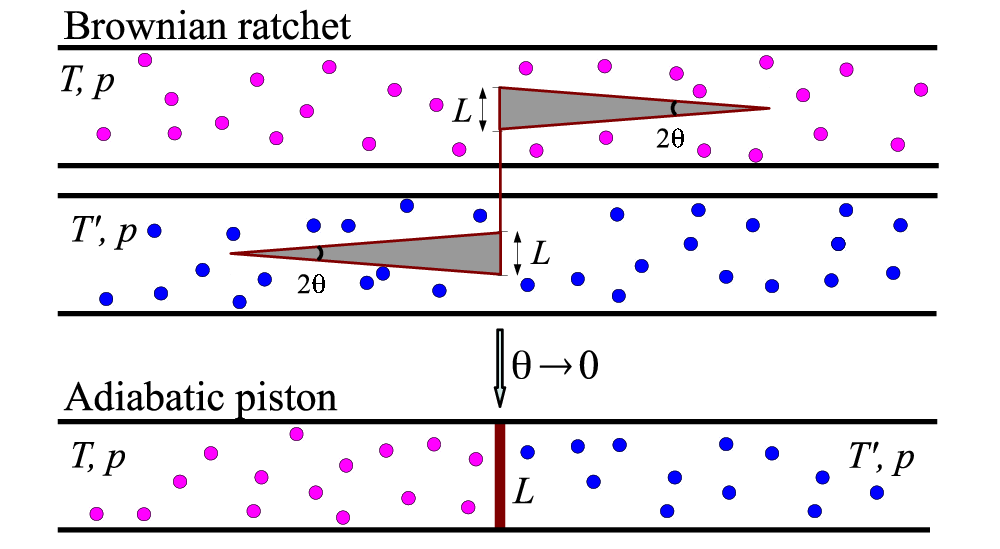}}
\caption{
A Brownian ratchet proposed in \protect\cite{VdB-prl2004-ratchet,*VdB-njp2005-ratchet}(top)
consists of two triangles (total mass $M$) that translates along the horizontal axis. This model can be
mapped to the adiabatic piston (bottom)
in the limit, $\theta\to 0$.  
\label{fig:3}} 
\end{figure}

{\it Ratchet model in two gas environments}: 
Van den Broeck {\it et al.} \cite{VdB-prl2004-ratchet,*VdB-njp2005-ratchet} proposed and analyzed a series of Brownian ratchet models that move horizontally in contact with two ideal gas environments at different temperatures $T$ and $T'$. One typical example is shown in Fig.~\ref{fig:3}(top), where we assumed isobaric condition only to simplify the argument without loosing the essential point. Microscopic methods concluded that it moves steadily with the base of the triangle in hotter environment being ahead, i.e. leftwards if $T>T'$. Based on our principle, the origin of nonequilibrium force is essentially the same as the aforementioned adiabatic piston. Intuitively, if we look at only the bases of triangles, it already appears identical to the adiabatic piston, Fig.~\ref{fig:3}(bottom). In fact, the sides of the triangle receive more frequent collisions than on the base but with much less impact on the horizontal motion. We can rigorously show that in the limit of $\theta\to 0$ (see Fig.~\ref{fig:3}), the momentum transfer rate on the sides converges to the equilibrium force, $pL$, without fluctuation or frictional velocity dependence [16]. Therefore, in this limit, the effect of side surface vanishes and the same principle as the adiabatic piston determines the motion of the ratchet model.  The result agrees with 
their perturbative calculation\cite{VdB-prl2004-ratchet,*VdB-njp2005-ratchet}.

{\it Inelastic triangle}: Costantini {\it et al.} \cite{0295-5075-82-5-50008} studied a variant of 
above ratchet model using a single triangle but with inelastic surface of restitution constant $e$.
In this case, the net house-keeping component vanishes, as if the triangle is in a hydrostatic pressure, $(1+e)p/2$. On the other hand, the excess dissipation $J^{(e)}_{\rm diss, ex}$ on the side surfaces are less important than that on the base, in the way that the contribution by the side surfaces vanishes in the limit $\theta\to 0$. In this limit, the force balance with frictional drag $-\gamma \bar{V}$ and Eq.~(\ref{eq:Jeex}) yields
\eqn{\bar{V}=-\frac{c}
{M}\frac{1-e}{2v_{\rm th}}\kT.}
This result is identical to the one obtained by microscopic approach \cite{0295-5075-82-5-50008} to the lowest order in $1-e$, if we choose $c=\sqrt{\pi/8}$.

In summary we have introduced a unified theory on the generation of nonequilibrium force as  momentum transfer deficit due to dissipation. This principle is applied to a new model of NESS, named, cooled/warmed piston, as well as to many existing models such as adiabatic piston 
 in a unified manner.
What we clarified here is that, while the energetics at Langevin level \cite{LNP} is enough to treat the dissipation,   the dissipation attributed to Brownian motion plays a decisive role \cite{feynman63-ch39} in the force generation  through the {MDD}. 
As perspectives, the {MDD} should be taken incorporated in the hydrodynamic description of adiabatic piston \cite{hydro-adiab-piston-malekmansour-PRE06}.  
It is of interest to generalize the present results to interacting gas particles, for examples the boundary thermostats \cite{boundary-thermostat-Hoover1989} as well as to the contact value theorem \cite{Henderson1979315,*10.1063/1.442238} under nonequilibrium.

We thank the members of Physico-Chimie Th\'eorique at E.S.P.C.I. 
K.S. thanks RIKEN for a financial support 
under the contract, CNRS: 30020830.
We acknowledge Hal Tasaki for the argument for $T_{\rm kin}$ of inelastic collision.

\bibliographystyle{apsrev4-1.bst}   
\bibliography{bib-AF}

\end{document}